\documentstyle[prl,aps,preprint,tighten,graphicx]{revtex}
\begin{document}
%\twocolumn[\hsize\textwidth\columnwidth\hsize
\draft
%\onecolumn[\hsize\textwidth\columnwidth\hsize
           \csname @twocolumnfalse\endcsname
%\preprint{\parbox{5cm}{MPG--VT--UR 203/00}} 
%\begin{titlepage}
\title{Equation of State for Strange Quark Matter in a Separable Model}
\author{C. Gocke}
\address{Fachbereich Physik, Universit\"at Rostock, D-18051 Rostock, Germany}
\author{D.~Blaschke}
\address{Institute for Nuclear Theory, University of Washington, Box 351550,
Seattle, WA 98195\\
Fachbereich Physik, Universit\"at Rostock, D-18051 Rostock, Germany\\
Bogoliubov Laboratory for Theoretical Physics, JINR, RU-141980 Dubna, Russia} 
\author{A. Khalatyan}
\address{Department of Physics, Yerevan State University, 375025 Yerevan, 
Armenia}
\author{H. Grigorian}
\address{Fachbereich Physik, Universit\"at Rostock, D-18051 Rostock, Germany\\
Department of Physics, Yerevan State University, 375025 Yerevan, Armenia}  
\maketitle
\begin{abstract}
We present the thermodynamics of a nonlocal chiral quark model with separable 
4-fermion interaction for the case of $U(3)$ flavor symmetry within a 
functional integral approach. 
The four free parameters of the model are fixed by the chiral condensate, and 
by the pseudoscalar meson properties (pion mass, kaon mass, pion decay 
constant).
We discuss the $T=0$ equation of state (EoS) which describes quark confinement 
(zero quark matter pressure) below the critical chemical potential 
$\mu_c=333$ MeV. The new result of the present approach is that the 
strange quark deconfinement is separated from the light quark one and occurs
only at a higher chemical potential of $\mu_{c,s}=492$ MeV.  
We compare the resulting EoS to bag model ones for two and three quark flavors,
which have the phase transition to the vacuum with zero pressure also at 
$\mu_c$.

We study quark matter stars in general relativity theory assuming 
$\beta$-equilibrium with electrons and show that for configurations  
with masses close to the maximum of stability at $M=1.62 \div 1.64 ~M_{\odot}$
strange quark matter can occur.   
\end{abstract}
%\vskip1cm]
%\date{\today}

\newpage
\section{Introduction}
Since the discovery of the parton substructure of nucleons and its 
interpretation within the constituent quark model, much effort has been spent 
to explain the properties of these particles. 
The phenomenon of {confinement}, i.e. the property of quarks to exist only 
in bound states as {mesons} and {baryons} in all known systems, poses 
great difficulties for a describing theory. 
So far the problem has been solved by introducing a {color} interaction 
that binds all colored particles to ``colorless'' states.  

However, it is believed and new experimental results \cite{cernlb} underline 
it, that at very high temperatures exceeding $150$ MeV, or densities higher 
than three times nuclear matter density, a transition to deconfined 
quark matter can occur. Besides of the heavy ion collisions performed in
particle physics, the existence of a {deconfinement phase} and 
its properties is of high importance for the understanding of 
{compact stars} \cite{glenden} in astrophysics. 
These, that are popular as Neutron stars, imply core densities above 
three times the nuclear saturation density \cite{nsdata} so that quark matter 
is expected to occur in their interior \cite{nsi} and several suggestions have 
been made in order to detect signals of the deconfinement transition 
\cite{gpw,pgb}.

Unfortunately, rigorous solutions of the fundamental theory of color 
interactions ({Quantumchromodynamics (QCD)}) for the EoS at finite baryon 
density could not be obtained yet, even Lattice gauge theory simulations have
serious problems in this domain \cite{udq}. 
To describe interacting quark matter it is therefore necessary to find 
approximating models. The best studied
one is the {Nambu-Jona-Lasinio (NJL)} model that was first developed to 
describe the interaction of nucleons \cite{njl1} and has later been applied
for modeling low-energy QCD \cite{volkov,sandy,hatsuda} with particular 
emphasis on the dynamical breaking of chiral symmetry and the occurence of 
the pion as a quasi Goldstone boson.  
The application of the NJL model for studies of quark matter thermodynamics is 
problematic since it has no confinement and free quarks appear well below the 
chiral phase transition \cite{klevansky,brunodis}. This contradicts to 
results from lattice gauge theory simulations of QCD thermodynamics where
the critical temperatures for deconfinement and chiral restoration coincide. 
That can be helped by using a separable model which can be treated similarly 
to the NJL model but includes a momentum dependence for the interaction via 
formfactors. It has been shown \cite{sepmod} that in the chiral limit the 
model has no free quarks below the chiral transition. 

Looking again at the densities of compact star cores and comparing with the 
results of NJL model calculations \cite{brunodis,buballa} it seems reasonable 
to include strange-flavor quarks in the model because the energy density is 
sufficiently high for their creation in weak processes. 
Therefore, we extend in the present work the separable model to the case of 
three quark flavors, assuming for simplicity $U(3)$ symmetry. 
We will calculate the partition function
using the method of bosonisation and applying the mean-field approximation. 
Finally we will formulate the resulting thermodynamics of three-flavor quark 
matter and obtain numerical results for the quark matter EoS and compact star 
structure.

%=============================================================================
% THEORY OF QUARK FIELDS
%=============================================================================
%\section*{Theory of Quark Fields}
\section{The separable quark model}

The starting point of our approach is an effective chiral quark model 
action with a four-fermion interaction in the current-current form 
%
%\newpage
\begin{eqnarray}
  \label{eq:4}
  {\cal S}[q, \bar q] &=& 
\int \frac{d^4\!k}{(2\pi)^4} \big[\bar q(k)i(\! \not k + \hat m)q(k)
              \nonumber \\
           &+& D_0 \int \frac{d^4\!k^\prime}{(2\pi)^4}\sum_{\alpha=0}^{8} 
                 [j_s^\alpha(k)j_s^\alpha(k^\prime) 
\nonumber \\
           \hspace{2cm} &+& j_p^\alpha(k)j_p^\alpha(k^\prime)]\big] ~, 
\end{eqnarray}
where we restrict us here to the {scalar} current 
$j_s^\alpha(k)=\bar q(k)\lambda_\alpha f(k)q(k)$
and the pseudoscalar current 
$j_p^\alpha(k)=\bar q(k)i\gamma_5\lambda_\alpha f(k)q(k)$
in Dirac space with $q(k)$ and $\bar q(k)$ being quark spinors and the 
formfactor $f(k)$ accounts for the nonlocality of the interaction.
The action (\ref{eq:4}) is invariant under chiral rotations of the quark 
fields and color correlations are neglected (global color model).
The generalization of previous models of this type \cite{sepmod,separable,bkt} 
to the three-flavor 
case is done by using the $U(3)$ symmetry  where $\lambda_\alpha$ are the 
Gell-Mann matrices and $\lambda_0=\sqrt{\frac{2}{3}}I$. 

Furthermore we do not include one of the possible models to account for the 
$U_A(1)$ anomaly since, at least in the quark representation of the Di Vecchia
- Veneziano model \cite{vv}, it can be shown that there is no contribution
to the quark thermodynamics on the mean-field level \cite{alkofer}.
For the quark mass matrix in flavor space we use the notation 
\begin{equation}
\label{hat}
\hat m=\sum_f m_f P_f~,
\end{equation}
where $m_f$ are the current quark masses and the projectors $P_f$ on the 
flavor eigenstate $f=u,d,s$ are defined as
\begin{eqnarray}\label{project} 
P_u&=&\frac{1}{\sqrt{6}}\lambda_0+\frac{1}{2}\lambda_3
+\frac{1}{2\sqrt{3}}\lambda_8,  \\
P_d&=&\frac{1}{\sqrt{6}}\lambda_0-\frac{1}{2}\lambda_3
+\frac{1}{2\sqrt{3}}\lambda_8,  \\
P_s&=&\frac{1}{\sqrt{6}}\lambda_0-\frac{1}{\sqrt{3}}\lambda_8~~.  
\end{eqnarray}
Since the Matsubara frequencies in the $T\to 0$ limit become quasicontinuous
variables, the summation over the fourth component $k_4$ of the 4-momentum 
has been replaced by the corresponding integration. 
According to the Matsubara formalism the calculations are performed 
in Euclidean space rather than in Minkowski space where we use 
$\gamma^4= i\gamma^0$.
The partition function in Feynman's path integral representation is given by
\begin{equation}
  \label{eq:5}
  {\cal Z}[T,\hat\mu] = \int {\cal D}\bar q {\cal D}q \exp\left(
         {\cal S}[q, \bar q]  - \int\frac{d^4\!k}{(2\pi)^4} 
 i\hat\mu\gamma^4\bar qq \right)~,
\end{equation}
where the constraint of baryon number conservation is realized by the diagonal
matrix of chemical potentials $\hat\mu$ (Lagrange multipliers) using the 
notation of the hat symbol analogous to (\ref{hat}).

In order to perform the functional integrations over the quark fields 
$\bar q$ and $q$ we use the formalism of bosonisation (see \cite{ripka} and 
references therein) which is based on the Hubbard-Stratonovich transformation 
of the four-fermion interaction terms employing the identity
\begin{eqnarray}
\exp\left\{ D_0 \int_k \int_{k^\prime} \sum_{\alpha=0}^{8} 
j_s^\alpha (k)j_s^\alpha (k^\prime)\right\}
= \nonumber \\
 {\cal N} \prod_\alpha\int d\sigma^\alpha 
\exp\left[\frac{(\sigma^\alpha)^2}{4 D_0} + 
\int_k \int_{k^\prime} j_s^\alpha(k) \sigma^\alpha(k^\prime)\right]~, 
\end{eqnarray}
for the scalar and a similar one for the pseudoscalar channel, where for the 
phase space integral the abbreviation $\int_k=\int\frac{d^4\!k}{(2\pi)^4}$ 
has been used and $\cal{N}$ is a normalization factor.
Now the generating functional is Gaussian in the quark fields and 
can be evaluated.
We arrive at the transformed generating functional in terms of bosonic 
variables 
\begin{eqnarray}  \label{eq:6}
{\cal Z}[T,\hat\mu] &=& \int\!\!\!\int{\cal D}\sigma^\alpha{\cal D}\pi^\alpha 
   \exp\left\{\cal{S}[\sigma^\alpha,\pi^\alpha]\right\}
\end{eqnarray}
with the action functional
\begin{eqnarray}
  \label{eq:7}
  {\cal S}[\sigma^\alpha,\pi^\alpha]=
&-&\int\frac{d^4\!k}{(2\pi)^4} \ln \left( \det_{DFC}  
              [\not{\tilde k} + \hat{m} +
              \hat{\sigma} f(\tilde k) + i\gamma_5\hat{\pi} f(\tilde k)] \right) \nonumber \\ 
&+& \frac{\bar\sigma_\alpha\bar\sigma^\alpha}{4D_0} + \frac{\bar\pi_\alpha\bar\pi^\alpha}{4D_0}~. 
\end{eqnarray}
with analogous use of the already known hat symbol and
the 4-vector $\tilde k_f = \left({\vec k},{k_4 +  i\mu_f} \right)$. 
In order to further evaluate the integral over the auxiliary bosonic 
fields $\sigma_\alpha$ and $\pi_\alpha$ we expanded them
around their mean values $\bar\sigma_\alpha$ and $\bar\pi_\alpha$ that 
minimize the action
\[\begin{array}{r@{\:=\:}l}
  \sigma_\alpha & \bar\sigma_\alpha + \tilde\sigma_\alpha(k) \\
  \pi_\alpha & \bar\pi_\alpha + \tilde\pi_\alpha(k) 
\end{array}\]
and neglect the fluctuations $\tilde\sigma_\alpha(k)$ and 
$\tilde\pi_\alpha(k)$ in the following. 
The mean values of the pseudoscalar field vanish for symmetry reasons 
\cite{alkofer}.
The indices $DFC$ refer to the determinant in Dirac-, flavor- and color-
space. So we end up with the mean-field action 
\begin{eqnarray}
  \label{eq:8}
  {\cal S}_{\rm{MF}}[T,\{\mu_f\}] &=& 
         \sum_f\left(-2N_c \int\frac{d^4\!k}{(2\pi)^4}[
             \ln(\tilde k_f^2 + M_f^2) ] + \frac{\Delta_f^2}{8D_0}\right)~,
\end{eqnarray}
with the effective quark masses 
$M_f=M_f(\tilde k)=m_f + \Delta_f f(\tilde k)$ and the number of colors $N_c$. 
The flavor dependent mass gaps $\Delta_f$ are defined by 
$\hat\sigma=\sum_f\Delta_fP_f$. 
\subsection{Quark matter thermodynamics in mean field approximation}
In the mean field approximation, the grand canonical thermodynamical potential
is given by
\begin{eqnarray}
  \label{eq:9}
\Omega(T,\{\mu\}) &=& 
\beta^{-1}\ln\left\{{\cal Z}[T,\{\mu_f\}]/{\cal
    Z}[0,\{0\}]\right\}\nonumber \\ 
&=& \beta^{-1} \left\{{\cal S}_{\rm{MF}}[T,\{\mu_f\}]-
{\cal S}_{\rm{MF}}[0,\{0\}]\right\}~,
\end{eqnarray}
Where the divergent vacuum contribution has been subtracted.
In what follows we consider the case $T=0$ only. 
In order to interpret our result, we want to represent it as a sum of three 
terms
\begin{eqnarray}
  \label{eq:12}
  \Omega(0,\{\mu\}) &=& \sum_f\left( -2N_c \int\frac{d^4\!k}{(2\pi)^4}
          \ln\left(\frac{\tilde k_f^2 + M_f^2}{\tilde k_f^2 + m_f^2}\right) 
                  + \frac{\Delta_f^2}{8D_0}\right) \nonumber \\
               && -2N_c \sum_f\left(\int\frac{d^4\!k}{(2\pi)^4}
                  \ln\left(\frac{\tilde k_f^2 + m_f^2}{k_f^2 + m_f^2}\right)\right) \\
               && +\sum_f\left( 2N_c \int\frac{d^4\!k}{(2\pi)^4}
                  \ln\left(\frac{k^2 + (M_f^0)^2}{k^2 + m_f^2}\right) 
                  - \frac{(\Delta_f^0)^2}{8D_0}\right)~, \nonumber
\end{eqnarray}
where $M_f^0=m_f+\Delta_f^0f(k^2)$ are the effective quark masses in the vacuum.
The second term on the r.h.s. of this equation is the renormalized 
thermodynamical potential of an ideal fermion gas \cite{kapusta}.
The third term of Eq.\ (\ref{eq:12}) is independent of $T$ and $\mu$, i.e.\ 
it is a (thermodynamical) constant for the chosen model. Refering to the MIT bag model 
we call this term the {bag-constant} $B$. 
The remaining term includes the effects of quark interactions in the mean field
approximation and can be evaluated numerically.

All thermodynamical quantities can now be derived from Eq.\ (\ref{eq:12}). 
For instance,
pressure, density, energy density and the chiral condensate are given by:
\begin{equation}
  \label{eq:13}
  pV = -\Omega\;,\;n = -\frac{\partial\Omega}{\partial\mu}\;,\;
  \varepsilon = -p+\mu n\;,\;
<\!\bar q_fq_f\!> = \frac{\partial\Omega}{\partial m_q}~.
\end{equation}

Still the quark mass gaps $\Delta_f$ have to be determined. This is done by 
solving the gap equations which follow from the minimization conditions
$\frac{\partial\Omega}{\partial\Delta_f}=0$. 
The gap equations read
\begin{equation}
  \label{eq:14}
  \Delta_f = 4D_0(-2N_c)\int\frac{d^4\!k}{(2\pi)^4}\frac{2M_ff(\tilde k)}
{\tilde k_f^2+M_f^2}~.
\end{equation}
As can be seen from (\ref{eq:14}), for the chiral $U(3)$ quark model 
the three gap equations for $\Delta_u,~\Delta_d,~\Delta_s$ are decoupled
and can be solved separately. 
%
%===========================================================
% NUMERICAL RESULTS
%===========================================================
\section{Results for the Gaussian formfactor}
\subsection{Parametrization of the model}
In the nonlocal separable quark model described above the formfactor of the interaction was not yet specified.
In the following numerical investigations we will employ a simple Gaussian 
\begin{equation}
f(k)=\exp(-k^2/\Lambda^2)~,
\end{equation}
which has been used previously for the description of meson \cite{birse} and 
baryon \cite{golli} properties in the vacuum as well as for those of 
deconfinement and mesons at finite temperature \cite{bkt,bbkmt}. 
A systematic extension to other choices of formfactors can be found in
\cite{scoccola,bubsepmod}.
 
The Gaussian model has five free parameters to be defined: the coupling  
constant $D_0$, the interaction range $\Lambda$, and the three current quark
masses $m_u$, $m_d$, $m_s$. Setting $m_u=m_d=:m_q$ we restrict ourselves to 
four free parameters.
These are fixed by the three well known observables:  
pion mass $m_\pi=140$ MeV, kaon mass $m_K=494$ MeV and pion decay 
constant $f_\pi=93$ MeV. 
The formulas for the meson masses and the decay constant 
are calculated as approximations of the {Bethe-Salpeter equation} 
including the generalized {Goldberger-Treiman relation} \cite{bubsepmod}.

The fourth condition comes from values
for the chiral condensate that are conform with phenomenology. 
The resulting parametrisations of the quark model are shown in 
Tab.\ \ref{tbl:1}.

\subsection{Thermodynamics for quark matter without $\beta$ equilibrium}
\begin{figure}[bth]
  \begin{center}
    \includegraphics[width=0.6\linewidth]{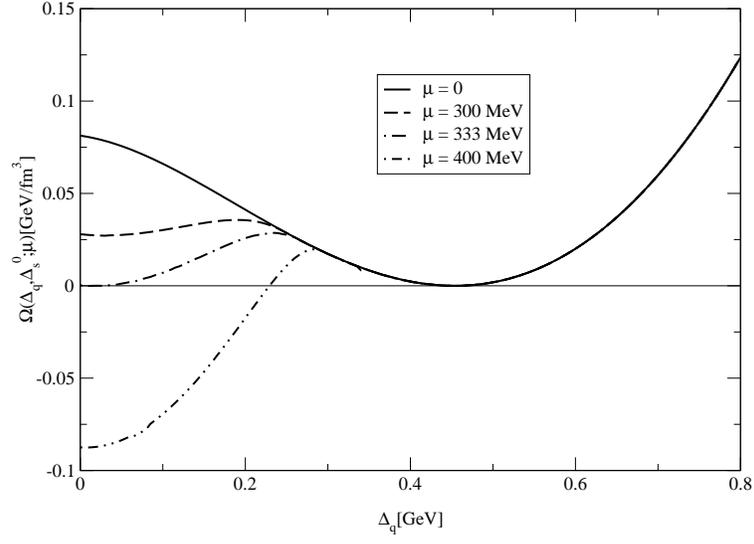} 
    \vspace*{0.2cm}
    \caption{Dependence of the thermodynamical potential on the light flavor 
        gap $\Delta_q=\Delta_u=\Delta_d$
        (order parameter) for different values of the chemical potential,
        $\Delta_s=682$ MeV.
        }
    \label{fig1}
  \end{center}
\end{figure}

\begin{figure}[bth]
  \begin{center}
    \includegraphics[width=0.6\linewidth]{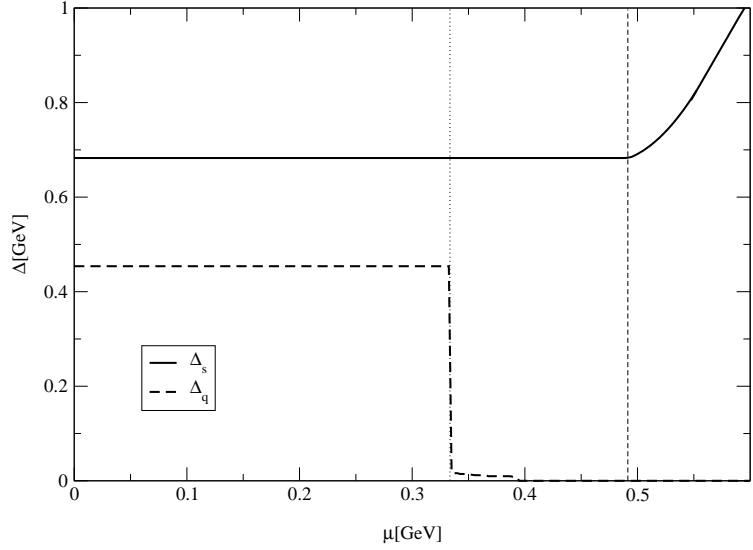} 
    \vspace*{0.3cm}
    \caption{Solutions of the gap equations that minimize the potential}
    \label{fig2}
  \end{center}
\end{figure}

This case is relevant for systems which are considered for time scales larger 
than the typical strong interaction time of about $1$ fm/c but smaller than 
the weak interaction time of several minutes, so that the presence of leptons  
(electrons) does not influence on the composition of quark matter and we can 
choose the chemical equilibrium with $\mu_u=\mu_d=\mu_s=\mu$.
For the numerical calculations we choose the parameter set for the light quark
condensate $-\langle\bar uu + \bar dd\rangle^{1/3}=240$ MeV which is a typical
value known from phenomenology. 
We consider the behavior of thermodynamical quantities at $T=0$ with 
respect to the chemical potential. 
As we set $m_u=m_d$ earlier there is no difference between up- 
and down quarks and both are referred to as light quarks.  
%\newpage
%
Fig.\ \ref{fig1} visualizes the behavior of the thermodynamical potential as a function of the light quark gap $\Delta_q=\Delta_u=\Delta_d$ for different 
values of the chemical potential $\mu$. 
For $\mu<\mu_c=333$ MeV the argument and the value of the global minimum is 
independent of $\mu$ which corresponds to a vanishing quark density 
(confinement). 
At the critical value $\mu=\mu_c=333$ MeV a phase transition occurs from the 
massive, confining phase to a deconfining phase negligibly small mass gap. 
From the solution of the gap equation shown in Fig. \ref{fig2}
one can see that the strange quark gap of $\Delta_s=682$ MeV still remains 
unchanged. Thus the strange quarks are confined until a higher value 
of the chemical potential $\mu_{c,s}=492$ MeV is reached. This value is much 
bigger than the current strange quark mass. 
In Fig.\ \ref{fig2} we separate by vertical lines the regions 
of full confinement, two-flavor deconfinement and full deconfinement.
Thus, in the present model, the onset of strange quark deconfinement is 
inhibited. Moreover, the onset of a finite strange quark density is not 
determined by a drop in the strange gap which remains constant and even 
starts to rise for large $\mu$ values.
This result of the present model drastically differs from those of 
bag models or NJL models.
The reason is the 4-momentum dependence of the dynamical quark mass function
which results in complex mass poles for the quark propagators and makes  
the naive identification of the mass gap with a real mass pole impossible
\cite{bubsepmod}.  

\begin{figure}[bth]
  \begin{center}
    \includegraphics[width=0.6\linewidth]{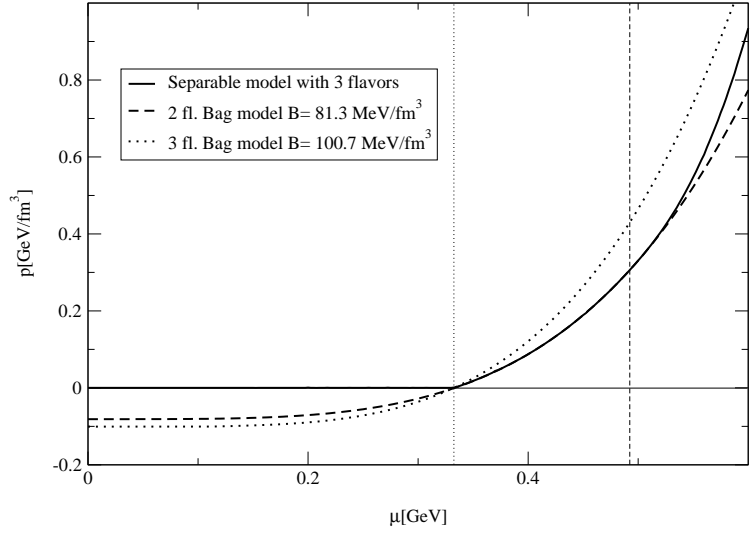} 
    \vspace*{0.3cm}
    \caption{Pressure of the quark matter as a function of the chemical 
potential for the separable model (solid line) compared to a three-flavor 
(dotted line) and a two-flavor (dashed line) bag model. All models have the
same critical chemical potential $\mu_c=333$ MeV for (light) quark 
deconfinement.}
    \label{fig3}
  \end{center}
\end{figure}

The effect on thermodynamical quantities can be understood if we look at the 
pressure. 
In Fig.\ \ref{fig3} we show for comparison the resulting equation of state for 
the pressure of the present separable model together with a two-flavor and a 
3-flavor bag model. Both bag models are chosen such that the critical chemical 
potential for the deconfinement coincides with that of the separable model.
The pressure of the present three-flavor separable model can be well described
by a two-flavor bag model with a bag constant $B=81.3$ MeV/fm$^3$ in the 
region of chemical potentials $333$ MeV $\le \mu \le 492$ MeV where the third 
flavor is still confined.
For comparison, the 3 flavor bag model has a bag constant $B=100.7$
MeV/fm$^3$ and is considerably harder than the separable one due to the 
additional relatively light strange quark flavor. 

\subsection{Inclusion of $\beta$ equilibrium with electrons}
\begin{figure}[bth]
  \begin{center}
    \includegraphics[width=0.6\linewidth]{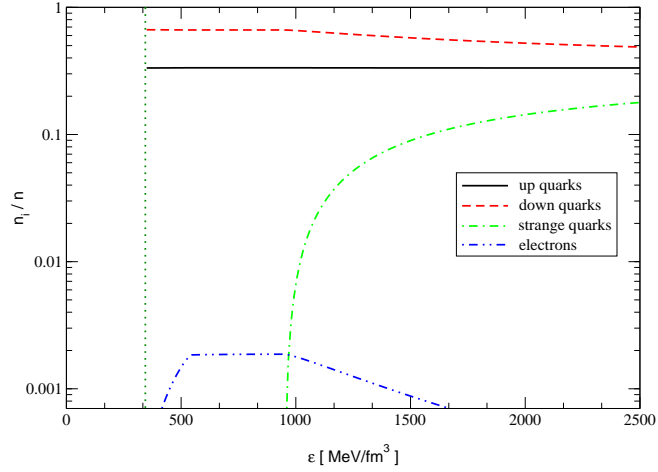} 
    \vspace*{0.3cm}
    \caption{Composition of three-flavor quark matter in $\beta$ equilibrium 
        with electrons.}
    \label{fig4}
  \end{center}
\end{figure}

\begin{figure}[bth]
  \begin{center}
\vspace{1cm}

    \includegraphics[width=0.6\linewidth]{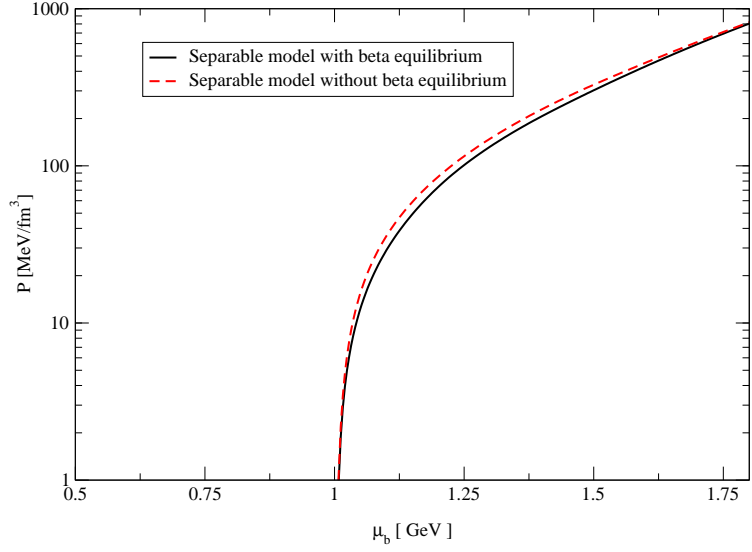} 
    \vspace*{0.3cm}
    \caption{Pressure of three-flavor quark matter with $\beta$ equilibrium 
        and without.}
    \label{fig5}
  \end{center}
\end{figure}

Quark matter in $\beta$-equilibrium is to be supplemented with the two 
relations for conservation of baryon charge and electric charge. 
In the deconfined phase there are quarks and leptons (in our model case up, 
down, strange quarks and electrons) with vanishing net electric charge 
\begin{eqnarray}\label{neutral}
Q_{{\rm q}}(\mu^u,\mu^d,\mu^s) +Q_{{\rm L}}(\mu^e) = {\frac{{2}}{{3}}} n_u -
{\frac{{1}}{{3}}}( n_d + n_s) - n_e = 0~.
\end{eqnarray}
Taking into account the energy balance in weak interactions
\begin{eqnarray}
d &\leftrightarrow& u + e^- +\bar \nu_e\\
s &\leftrightarrow& u + e^- +\bar \nu_e
\end{eqnarray}
and introducing the average quark chemical potential  
$\mu=\frac{1}{3}(\mu_u+\mu_d+\mu_s)$
we can write the $\beta$ equilibrium conditions as
\begin{eqnarray}\label{beta}
\mu_u =  \mu - {\frac{{2}}{{3}}} \mu^e\,
,\qquad \mu_d = \mu_s = \mu + {\frac{{1}}{{3}}} \, \mu^e \ .
\end{eqnarray}
Solving the equation of charge neutrality (\ref{neutral}) one can find the
chemical potential of electrons as a function of $\mu$ and using Eqs. 
(\ref{beta}) the equation of state can be given in terms of a single chemical 
potential $\mu$. 
In Fig.~\ref{fig4} we show the composition of the three-flavor quark matter
for the Gaussian separable model in the case of $\beta$ equilibrium with 
electrons as a function of the energy density.
As for the case without $\beta $ equilibrium we can define also in Fig. 
\ref{fig4} the regions of quark confinement 
($\varepsilon<\varepsilon_c=350$ MeV/fm$^3$) and three-flavor deconfinement 
($\varepsilon>\varepsilon_{c,s}=930$ MeV/fm$^3$). 
In the region of two-flavor deconfinement the concentrations of electrons,
up- and down- quarks coincide with those of the two-flavor bag model except 
for the relatively small energies close to $\varepsilon_c$ where the effect of 
a small dynamical quark mass leads to a density dependence of the composition.
In Fig. \ref{fig5} we demonstrate the influence of the $\beta$ equilibrium on 
the equation of state. It can be seen that the difference between pressures 
with and without  $\beta$ equilibrium is limited to the region of intermediate 
densities, where the electron fraction reaches its maximum value 
$x_e\simeq 0.002$.

\section{Applications for compact stars}

One of the main goals for studying the strange quark matter equation of state
is the possible application for compact stars. In particular, the hypothesis 
that strange quark matter might be more stable than ordinary nuclear matter 
\cite{strange} has lead to the investigation of possible consequences for 
properties of compact stars made thereof \cite{sstar}.
Most of these applications use the bag model equation of state where the 
result depends on the value of the bag constant as a free parameter.
Recently, first steps have been made towards a description of strange quark 
matter within dynamical quark models such as the NJL model \cite{snjl}, 
where the parameters are fixed from hadron properties. 
The non-confining quark dynamics of this model, however, leads to
predictions for dynamical quark masses and critical parameters of the
chiral phase transition which differ from those of confining models
\cite{mnstar,sepmod} and might be quantitatively incorrect.
Here we want to extend previous studies of compact star properties with 
dynamically confining quark models to the strange quark 
sector and find the characteristics of stable compact star configurations 
with the equation of state derived above. 

For the calculation of the self-bound configuration for the quark matter with 
gravitational interaction one needs the condition of mechanical equilibrium of 
the thermodynamical pressure with the gravitational force.
This condition is given by the Tolman-Oppenheimer-Volkoff-Equation
\begin{eqnarray}
\frac{dP}{dr}=- G (\varepsilon(r) + P(r)) \frac{m(r) 
+ 4\pi G r^3 P(r)}{r(r-2Gm(r))}~
\end{eqnarray} 
and defines the profiles for all thermodynamical quantities in the case of nonrotating spherically symmetric distributed matter configurations in general relativity.
In this equation $m$ denotes the accumulated mass in the sphere with radius 
$r$ given by
\begin{equation}
m(r)=4 \pi \int_0^r \varepsilon(r') r'^2 dr
\end{equation}
The gravitational constant is denoted by $G$. 
The radius $R$ of the star is defined by the condition that the pressure 
becomes zero on the surface of the star $P(R)=0$.
The total mass of the star is $M=m(R)$.

\begin{figure}[bth]
  \begin{center}
    \includegraphics[width=0.6\linewidth,angle=-90]{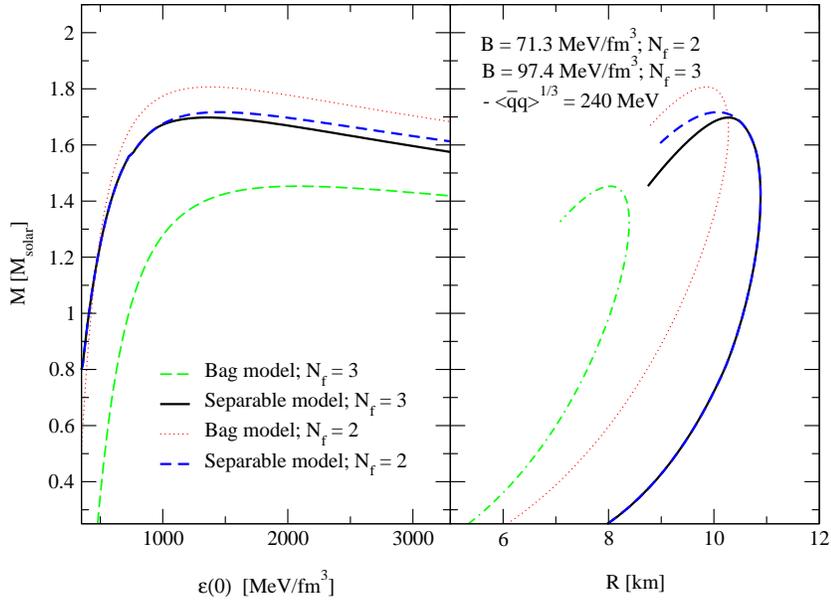} 
    \vspace*{0.3cm}
    \caption{Stability for compact stars composed of quark matter.}
    \label{fig6}
  \end{center}
\end{figure}

Each configuration has one independent parameter which could be chosen to be 
$\varepsilon(0)$, the central energy density. 
In Fig.~\ref{fig6} we show the dependence of the total mass of the 
configuration as a function of the central density and the radius for the 
separable quark model and for the bag model in the
cases of two and three flavors respectively. 

The rising branches of the mass-radius or mass-density relations correspond to 
the families of stable compact stars.
The maximum possible mass for the separable model is $1.64~M_\odot$ for the 
three-flavor and $1.71~M_\odot$ for the two-flavor case.
The maximal central density is about $1350$ MeV/fm$^3$ which allows for the 
three-flavor case to have strange quark matter in the core of the quark star. 
The comparison with the corresponding bag model strange stars shows that the 
latter are more compact, their maximum radius is about $8$ km, and less massive
with a maximum mass of about $1.5~M_\odot$.
The maximum radius of stars within the separable model is 11 km and thus 
exceeds the radii for both two and three flavor bag model quark stars. 
The origin
of this difference is the behavior of the pressure in the low density region.  

%==================================================================
%CONCLUSION
%==================================================================
\section{Conclusion}
For neutron stars it is relevant to include the effects of strange flavor in a 
model for quark matter. In the simplest case considering $U(3)$ symmetry this
can be done without increasing the complexity of the generating functional.  
We showed that in our separable model the gap equations decouple and can be 
solved separately. The resulting thermodynamics can be solved numerically and
gives the equations of state for interacting quark matter. Unlike the well
known NJL model the separable model is able to express the effects of 
confinement in the thermodynamical quantities.
The new result obtained within the present approach is the separation of the
deconfinement of light quark flavors at $\mu_{c}=333$ MeV from that of
strange quarks which occurs only at a higher chemical potential of 
$\mu_{c,s}=492$ MeV.
A consequence for the application of the EoS presented here in compact
star calculations is that strange quarks do occur only close to the maximum 
mass of $1.64~M_{\odot}$, i.e. that for masses below $1.62~M_{\odot}$,
only two-flavor quark matter can occur. 
\section*{Acknowledgments}
We like to thank our colleagues  M.\ Buballa, Y.\ Kalinovsky, M.\ Ruivo, 
N.\ Scoccola and P.C.\ Tandy. 
Their studies of separable quark models have helped us to 
formulate the present one. We are grateful to R. Alkofer for pointing out 
Ref. \cite{alkofer} to us.
D.B. thanks for partial support of the Department of 
Energy during the program INT-01-2: "Neutron Stars" at the 
University of Washington.
H.G. is grateful to DFG for support under grant number 436 ARM 17/5/01.
We acknowledge the support of DAAD for scientist exchange between the 
Universities of Rostock and Yerevan.  
%

%%%%%%%%%%%%%%%%%%%%%%%%%%%

%
%\vspace*{-0.2cm}
%\noindent
%\begin{minipage}{\textwidth}
\begin{table}[bth]
  \begin{tabular}{c||c|c|c|c||c|c}
    \hline
 $-<\!\bar uu + \bar dd\!>^{1/3}$ & $\Lambda$ & $D_0$ & $m_q$ & $m_s$ & $\Delta_q^0$ & $\Delta_s^0$ \\
{[MeV]}&[MeV]&[GeV$^{-2}$]&[MeV]&[MeV]&[MeV]&[MeV]\\
    \hline
    $230$ & $659.2$ & $29.32$ & $6.8$ & $143.5$ & $549.9$ & $767.8$\\
    $235$ & $697.6$ & $23.88$ & $6.4$ & $136.1$ & $497.0$ & $719.3$\\   
    $240$ & $736.5$ & $19.88$ & $6.0$ & $129.5$ & $453.8$ & $682.1$\\   
    $245$ & $775.5$ & $16.85$ & $5.6$ & $123.4$ & $419.7$ & $653.1$\\   
    $250$ & $814.9$ & $14.49$ & $5.3$ & $118.0$ & $391.1$ & $630.0$\\   
    $255$ & $853.8$ & $12.66$ & $5.0$ & $112.9$ & $368.7$ & $611.4$\\   
    $260$ & $894.2$ & $11.11$ & $4.7$ & $108.1$ & $349.1$ & $596.1$\\ 
    \hline
  \end{tabular}
  \vspace*{0.2cm}
  \caption{Parameter sets for the Gaussian separable model for different values of the chiral condensate $<\!\bar uu + \bar dd\!>$. }\label{tbl:1}
\end{table}
%\end{minipage}

\vspace*{-0.2cm}

 
\end{document}